\documentclass[aps,pre,preprint,groupedaddress,showpacs]{revtex4}
\usepackage{latexsym,amsmath,amssymb,amsthm,amsfonts,graphicx}

\newcommand{\nc}{\newcommand} 
\nc{\diff}[2]{\frac{d #1}{d #2}}
\nc{\diffn}[3]{\frac{d^{#3} #1}{d {#2}^{#3}}} 
\nc{\abs}[1]{\lvert #1 \rvert} 
\DeclareMathOperator{\sech}{sech}
\DeclareMathOperator{\G}{\Gamma}
\nc{\intinf}{\int_{-\infty}^{\infty}}
\nc{\cE}{{\mathcal E}}
\nc{\cL}{{\mathcal L}}
\nc{\cW}{{\mathcal{W}}}
\nc{\vc}{v_{\rm c}}
\nc{\vin}{v_{\rm in}}
\nc{\vout}{v_{\rm out}}
\nc{\Vc}{V_{\rm c}}
\nc{\Vin}{V_{\rm in}}
\nc{\Vout}{V_{\rm out}}
\nc{\half}{\frac{1}{2}}
\nc{\Xs}{X_{\rm S}}
\nc{\Xmax}{X_{\rm max}}

\begin{document}


\title{Vector-soliton collision dynamics in nonlinear optical fibers}

\author{Roy H. Goodman}
\email{goodman@njit.edu}
\homepage{http://math.njit.edu/goodman/}
\affiliation{Department of Mathematical Sciences, New Jersey Institute of
  Technology, Newark, NJ 07102}
\author{Richard Haberman}
\email{rhaberma@mail.smu.edu}
\affiliation{Department of Mathematics, Southern Methodist University, Dallas, 
TX 75275}

\date{\today}

\begin{abstract}
We consider the interactions of two identical, orthogonally polarized vector solitons in a nonlinear optical fiber with two polarization directions, described by a coupled pair of nonlinear Schr\"odinger equations.  We study a low-dimensional model system of Hamiltonian ordinary differential equations (ODEs) derived by Ueda and Kath and also studied by Tan and Yang.  We derive a further simplified model which has similar dynamics but is more amenable to analysis.  Sufficiently fast solitons move by each other without much interaction, but below a critical velocity the solitons may be captured. In certain bands of initial velocities the solitons are initially captured, but separate after passing each other twice, a phenomenon known as the two-bounce or two-pass resonance.  We derive an analytic formula for the critical velocity.  Using matched asymptotic expansions for separatrix crossing, we determine the location of  these ``resonance windows.''   Numerical simulations of the ODE models show they compare quite well with the asymptotic theory.  
\end{abstract}

\pacs{42.65.Tg, 05.45.Yv, 42.81.Dp, 05.45a}
\maketitle

\section{Introduction}
Solitary waves are an important phenomenon in nonlinear physics and applied mathematics.  Solitary waves have been studied in a diverse array of physical models including water waves~\cite{Russell,KdV,Ben:67}, quantum electronic devices (Josephson junctions)~\cite{BEMS:71}, and cosmology~\cite{AOM:91,BV:01}  One of the most important applications is to nonlinear optical communications where solitary waves have been proposed as information bits in optical fiber transmission systems~\cite{HT} and produced experimentally about 25 years ago~\cite{MSG:80}.  Other solitary wave phenomena in nonlinear optics include gap solitons in Bragg gratings~\cite{AW,CJ} and dispersion managed solitons~\cite{dmsx,dmnls}, which hold promise for eliminating the timing jitter associated with soliton transmission systems.

A single solitary wave propagating through a uniform medium appears particle-like in its coherence and steady propagation.  Of great interest are the interaction of multiple solitary waves and the behavior of solitary waves propagating through non-uniform media.  Solitary waves of completely integrable equations are known as solitons, and their interactions can be described completely, using multiple-soliton formulas derived via the inverse scattering transform~\cite{AS:81}.  The infinite set of conservation laws in integrable systems severely constrain the dynamics: collisions are elastic, and the solitons will re-emerge from a collision propagating with their initial amplitudes and speeds intact, although their positions will have undergone a finite jump.  Solitary wave collisions in non-integrable wave equations, can usually not be found in closed form, and show a much richer variety of behaviors: the waves may attract or repel each other and, upon collision, the solitary waves may lose their coherence and break apart, merge into a single localized structure, or even oscillate about one another.~\cite{KM:89,TY00,TY:01,TY:01a,CPS:86,CP:86,CSW:83,PC:83}.  

In a soliton-based communications system, the bits are represented by solitons.  In the simplest scenario, the presence of a soliton in a given timing window codes a one, and its absence codes a zero.  
Collisions between solitons, coupled with random noise in fiber characteristics, can lead to large perturbations in the solitons polarizations and to timing jitter~\cite{MGH:95}.  A bit that arrives at the wrong time may be interpreted incorrectly by a receiver, as would a soliton that splits in half or two solitons that merge. Ueda and Kath show such behaviors are possible and cite several additional numerical studies of solitons collisions not included here.  We an approach to the modeling and analysis of these phenomena that, while highly idealized, leads to new insights into these collisions.

Interacting pairs of solitary waves from several distinct (non-integrable) physical models have shown an interesting behavior in common.  At high speeds, the solitary waves move right past each other, hardly interacting, while at speeds below some critical velocity, the solitary waves interact strongly and may merge into a single localized state.  Interspersed among the initial velocities that lead to this capture are ``resonance windows'', for which the two waves approach each other, interact with each other for a finite time, and then move apart again; see the second and third graph in figure~\ref{fig:tanyang}. This has been explored by Tan and Yang in a system of coupled nonlinear Schr\"odinger (CNLS) equations that model nonlinear propagation of light in birefringent optical fibers~\cite{TY00,TY:01,TY:01a}, and by Cambell and collaborators in kink-antikink interaction in the $\phi^4$ equations and several other nonlinear Klein-Gordon models~\cite{CPS:86,CP:86,CSW:83,PC:83}.  These windows form a complicated fractal structure that has been described qualitatively and even quantitatively, but for which the underlying mechanism has been poorly understood.

The same phenomenon was also observed by Fei, Kivshar, and V\'azquez in the interaction of traveling kink solutions of the sine-Gordon and $\phi^4$ equations with weak localized defects~\cite{FKV:92a,FKV:92,KFV:91}.  Instead of two solitary waves merging, in this case the soliton could be captured, or pinned, at the location of the defect.  Almost all of the described models have been studied using the so-called variational approximation, in which the complex dynamics of the full PDE are modeled by a small, finite-dimensional system of ordinary differential equations.

The sine-Gordon equation with defect and the birefringent fiber-optic model discussed above feature a small parameter measuring the ``non-integrability'' of the system.  In a recent publication, Goodman and Haberman~\cite{GH:04}, exploited this small parameter to construct approximate solutions to the system of  ODEs  for the sine-Gordon model derived in~\cite{FKV:92}.  We calculated the critical velocity for defect-induced soliton capture via an energy calculation involving separatrix crossing, and the location of the resonance windows using a quantization condition that occurs in the asymptotic expansion.  In the current paper, we apply the same method to derive similar quantitative features in Ueda and Kath's ODE model of solitary wave collision in coupled nonlinear Schr\"odinger equations, and to explain the structure underlying the fractal structure of resonance windows.

In section~\ref{sec:theproblem}, we
introduce the physical model---a coupled system of Nonlinear Schr\"odinger
equations---and describe previous results in which the ``two pass resonance''
phenomenon has been observed. In section~\ref{sec:themodel}, we introduce Ueda
and Kath's finite-dimensional model system that captures the observed
dynamics, and introduce a simplified model which partially linearizes the
system and renders it amenable to our analysis.  We show numerically that this
simplification does not qualitatively alter the dynamics.  In section~\ref{sec:DeltaE},
we set up the calculation as a singular perturbation problem, and describe the
unperturbed dynamics.  We determine the critical velocity by calculating the
energy that is lost to vibrations as the solitons pass each other, employing a
Melnikov integral.  We generalize this calculation slightly for subsequent
interactions.  In section~\ref{sec:matching}, we construct approximate
solutions using matched asymptotic approximations, incorporating the previously calculated energy
changes.  Section~\ref{sec:nonlinearity} contains a
discussion of the differences between the original model and its
simplification and presents a weakly nonlinear theory to account for them.  We
conclude in section~\ref{sec:conclusion} with a physical summary and a discussion on
the applicability of these results to other systems displaying similar
behaviors. 
 
\section{Physical problem and prior results}
\label{sec:theproblem}
Following the previously cited~\cite{UK:90, TY:01}, we consider the model of polarized light propagation in a  optical fiber, given by the system of coupled nonlinear Schr\"odinger equations 
\begin{equation} 
\begin{split} 
\label{eq:CNLS}
 i \partial_t A + \partial_z^2 A + \left(\abs{A}^2 + \beta \abs{B}^2 \right) A &=0 \\
 i \partial_t B + \partial_z^2 B + \left(\abs{B}^2 + \beta \abs{A}^2 \right) B &=0.
\end{split}
\end{equation}
This system replaces the more familiar scalar Schr\"odinger equation when polarization is taken into effect~\cite{Agrawal}. The equations may be derived using the slowly varying envelope approximation
to Maxwell's equations in an optical fiber waveguide.  The variables  $A$ and $B$ describe the envelopes of wave
packets in the two polarization directions and $\beta$ is the nonlinear cross-phase modulation (XPM) coefficient that arises due to cubic ($\chi^{(3)}$) terms in the dielectric response of the glass.  Here we use $z$ as a space-like variable and $t$ as a time-like variable.  Of course, in the optics interpretation,
the labels $z$ and $t$ are switched, as the signal is defined as a function of
time at $z=0$ and the evolution occurs as the pulse moves down the length of
the fiber.  For mathematical simplicity, we will use $t$ as the evolution
variable. 

Our interest is in the interaction of solitary waves in the above system.  In the cases $\beta=0$ and $\beta=1$, system~\eqref{eq:CNLS} is completely integrable~\cite{ZS, Manakov}.  In the first case it reduces to a pair of uncoupled NLS equations; in the second it is known as the Manakov system.  For other values of $\beta$, the equations are not integrable.  Of special interest is the case $\beta= \frac{2}{3}$, which corresponds to linear fiber birefringence.  For very small values of $\beta$, this system models light propagation in a two-mode optical fiber~\cite{CDP}. In the case $\beta=0$, the equations are simply a pair of focusing nonlinear Schr\"odinger equations, with well known soliton solutions, first suggested as carriers of optical signals by Hasegawa and Tappert~\cite{HT}.  When $\beta$ takes any other value, the equations are non-integrable.  Yang~\cite{Yang97} studied these equations in great detail, enumerating several families of solitary waves and determining their stability.  Of these, the only stable solitary waves come from a family of symmetric single-humped solutions.  

The simplest solutions of interest to~\eqref{eq:CNLS} consist of an exponentially-localized soliton in the first component, $A$, and zero in the second component, $B$, or vice versa.  A single soliton propagates at constant speed with a fixed spatial profile.  An important problem is the interaction of two such solitons upon collision, as interactions between two such solitons may lead to errors in a soliton-based transmission system.

\begin{figure}
\begin{center}
\includegraphics[width=3in]{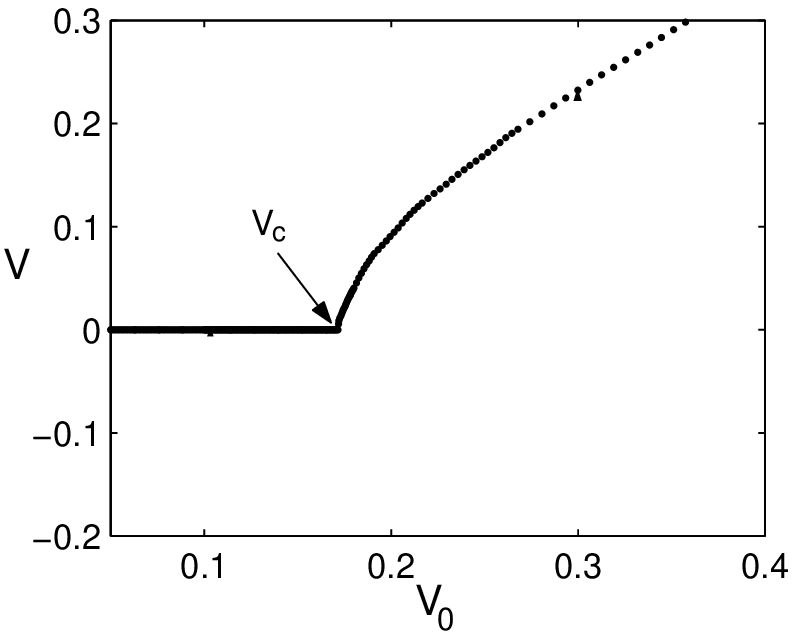}
\includegraphics[width=3in]{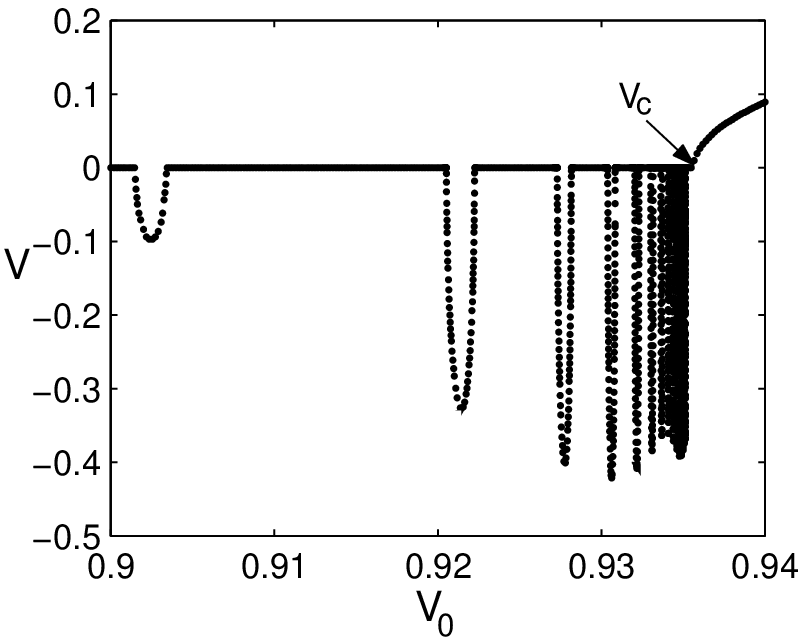}
\includegraphics[width=3in]{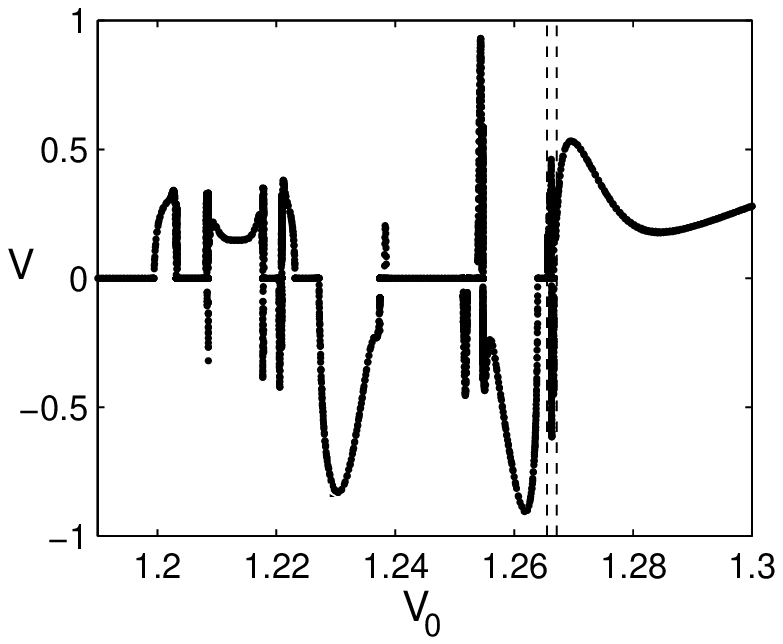}
\caption{The exit velocity as a function of the input velocity for $\beta=0.05$, $\beta=0.2$, and $\beta=0.6$, from Tan and Yang~\cite{TY:01}, original authors' annotations removed.}
\label{fig:tanyang}
\end{center}
\end{figure}

Tan and Yang numerically studied the interaction of two solitons initialized in orthogonal channels with identical amplitude, headed toward each other with exactly opposite initial velocities~\cite{TY00,TY:01,TY:01a}.  
For small values of $\beta \approx 0.05$, their simulations show that waves traveling above a critical velocity $\vc$, the solitons pass by each other, losing a little bit of speed, but not otherwise showing a complicated interaction.  At initial velocities below $\vc$, the solitons capture each other and merge into a stationary state near their point of collision.   See figure~\ref{fig:tanyang}, which shows three graphs from~\cite{TY:01} in which the input velocity of the solitons is plotted on the $x$-axis, and the solutions followed until they separate, and their outgoing velocity plotted on the $y$-axis, and assigned the value zero if they merge.

For somewhat larger values of $\beta \approx 0.2$, they find that in addition to the above behavior, that the capture region is interrupted by a sequence of ``resonant reflection windows.''  Solitons with initial velocities in these resonance windows are reflected instead of being captured.  The numerical simulations show that the solitons pass each other once, undergo a finite number of width oscillations, then pass each other a second time.  Thus they call this the ``two-pass'' resonance.  

For larger values of $\beta \approx 0.6$, they find not only reflection windows, but an intricate fractal-like structure of both reflection and transmission windows.  Certain portions of the structure, when properly scaled, look like copies, in come cases even reflected copies, of other portions of the structure, and such features are seen at many different scales.  

The two-bounce resonance in kink-antikink interactions was explained qualitatively in the first papers of the Campbell group~\cite{CSW:83,PC:83}.  As the kinks approach each other, they begin to interact,  and, at time $t_1$,  and energy is transferred into a secondary mode of vibration, with some characteristic frequency $\omega$.  If the initial velocity is below a critical value, the kinks no longer have enough energy to escape each other's orbit, and turn around to interact a second time $t_2$.  They show numerically that a resonant reflection occurs if $t_2 - t_1 \approx 2\pi n/\omega +\delta$. The parameter $\delta$ is found by a least squares fit with numerical data. This relation is used to estimate the resonant initial velocities.  This reasoning has subsequently been adapted in studies of sine-Gordon kink-defect interactions~\cite{FKV:92,KFV:91} and of vector soliton collisions~\cite{TY00,TY:01,TY:01a} which are the focus of this paper.

\section{The model equations}
\label{sec:themodel}
In order to gain further insight into the resonance phenomenon, Tan and Yang examine a model system
derived by Ueda and Kath~\cite{UK:90} using the variational method.  In the variational method, the solution is assumed to take a certain functional form $A(\vec p(t))$, $B(\vec p(t))$, dependent on parameters $\vec p(t)$ that are allowed to vary as a function of time.  This ansatz is then substituted into the Lagrangian functional for the PDE, which is integrated in space to yield a finite-dimensional effective Lagrangian,
$$
L_{\rm eff} = \int_{-\infty}^{\infty} \cL(A,A^*,B,B^*) dz,
$$
whose Euler-Lagrange equations describe the evolution of the time-dependent parameters.Equation~\eqref{eq:CNLS} has Lagrangian density.
\begin{equation}
\label{eq:CNLS_Laga}
\cL = i(A A_z^* - A_z A^*) + i( B B_z^* - B_z B^*) + (\abs{A_t}^2 -\abs{A}^4)+( \abs{B_t}^2-\abs{B}^4) -2 \beta\abs{A}^2\abs{B}^2
\end{equation}
Many examples using this method for PDE's arising as Euler-Lagrange equations are given in a recent review by Malomed~\cite{M:02}.

Following~\cite{UK:90}, we take an ansatz corresponding to two solitons at distance $2X$ of equal magnitude heading toward each other with equal speed, 
\begin{equation}
\begin{split}
\label{eq:ansatz}
A & = \eta \sech{\frac{z-X}{w}} \exp {i \left(v (z-X) + \frac{b}{2w}(z-X)^2 +\sigma\right)}, \\
B & = \eta \sech{\frac{z+X}{w}} \exp {i \left(-v (z+X) + \frac{b}{2w}(z+X)^2 +\sigma\right)}
\end{split}
\end{equation}
where $\eta$, $X$, $w$, $v$, $b$, and $\sigma$ are time-dependent parameters for the amplitude, position, width, velocity, chirp, and phase, whose evolution remains to be determined.  The variational procedure yields a conserved quantity $K = \eta^2 w$,
related to the conservation of the $L^2$ norm in CNLS, as well as the relations $\diff{X}{t}=v$ and $\diff{w}{t}=b$.
The evolution is described by the Euler-Lagrange equations
\begin{subequations}
\begin{align}
\diffn{X}{t}{2}&= \frac{16K\beta}{w^2}\frac{d}{d\alpha} F(\alpha)  \label{eq:TY1}\\
\diffn{w}{t}{2}&=\frac{16}{\pi^2 w^2}\left(\frac{1}{w}-K 
- 3 \beta K \frac{d}{d\alpha}\left(\alpha F(\alpha)\right)\right) \label{eq:TY2}
\end{align}
\label{eq:TanYang}
\end{subequations}
where $\alpha= X/w$ and the potential and coupling terms are given by 
\begin{equation}
\label{eq:F}
F(x) = \frac{x \cosh{x} - \sinh{x}}{\sinh^3{x}}.
\end{equation}
Note that $F$, actually $-F$, is a potential term, not a force.  We keep this notation for continuity with previous studies.

Numerical simulations show that for small $\beta$, a solution to~\eqref{eq:CNLS} with an initial condition of the form given in ansatz~\eqref{eq:ansatz} will remain close to that form, i.\ e.\ the solution will continue to consist of two nearly orthogonally polarized solitons, at least until they merge into a single bound state.  Using the symmetries of equation~\eqref{eq:TanYang}, we may set $K=1$ without loss of generality. Equivalently, the PDE symmetry may be used to set $K=1$ in the ansatz used by the variational method.

\begin{figure}
\begin{center}
\includegraphics[width=3in,angle=-90]{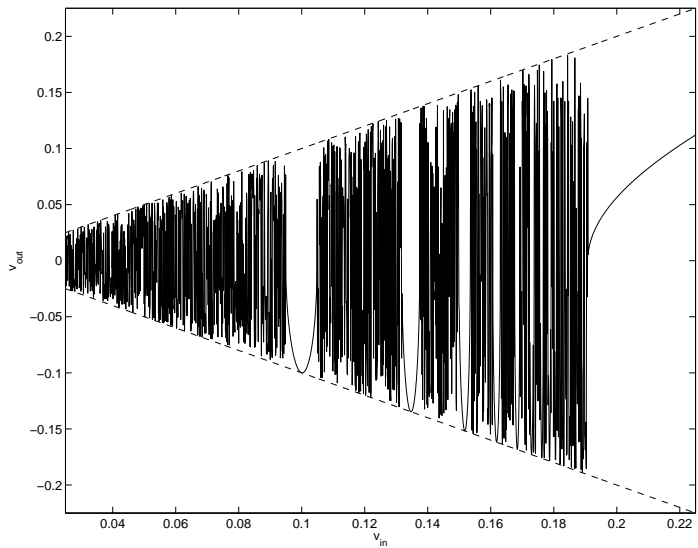}
\caption{The input vs.\ output velocity of a pair of orthogonally polarized solitons with $\beta=0.05$}
\label{fig:R003}
\end{center}
\end{figure}

These equations display the two-bounce resonance phenomenon, as shown by Tan and Yang.
Consider the initial value problem, with ``initial'' conditions describing the behavior as $t \to -\infty$.
$$
X \to -\infty;\; 
\diff{X}{t} \to \vin>0;\;
w \to 1; \; 
\diff{w}{t} \to 0
$$
This does not strictly determine a unique solution, since the solution is invariant to time translation.  We  plot $\vout$ as a function of $\vin$ with $\beta=0.05$ in figure~\ref{fig:R003}.  These and all other ODE simulations were performed using routines from ODEPACK~\cite{ODEPACK}.
Compare this figure to the three plots of figure~\ref{fig:tanyang}.  There are key similarities and differences between this graph and the exit velocity graphs of the full PDE simulations.  The critical velocity in this figure is about $\vc=0.19$, close to the value $\vc=0.1715$ found in~\cite{TY:01}.
A noteworthy difference is the complex behavior of solutions with initial velocity below $\vc$---no such behavior, not even the two-pass windows, was seen in the very careful simulations of Tan and Yang.  This should not be surprising, as system~\eqref{eq:TanYang} is Hamiltonian, and the set of initial conditions leading to unbounded trajectories in backwards time and bounded trajectories in forward time has measure zero, by reasoning similar to Poincar\'e recurrence, as shown in Proposition~1 of~\cite{GHW:02}.  Localized solutions to~\eqref{eq:CNLS} may lose energy to radiation modes, a dissipation mechanism not present in the ODE model.  As a further result of the dissipation, the output speeds of the reflected solutions are much smaller than the input speeds in the PDE solutions, whereas at the very center of the ODE windows, the output speed exactly matches the input speed.
A more interesting difference can be seen in the presence of the wide reflection windows, which were not found in the PDE simulations with this value of $\beta$, summarized in figure~\ref{fig:tanyang}.

\begin{figure}
\begin{center}
\includegraphics[width=3in]{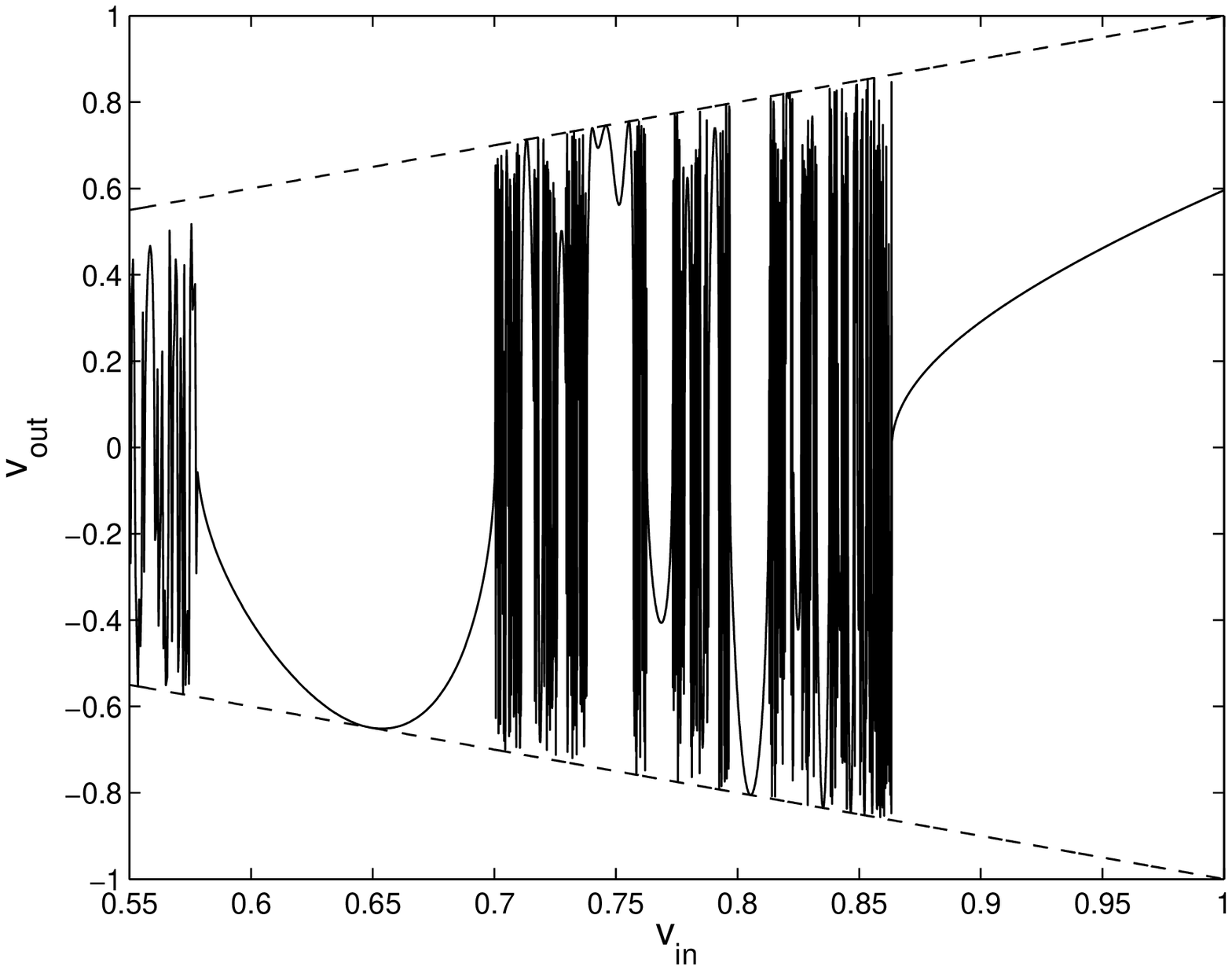}
\includegraphics[width=3in]{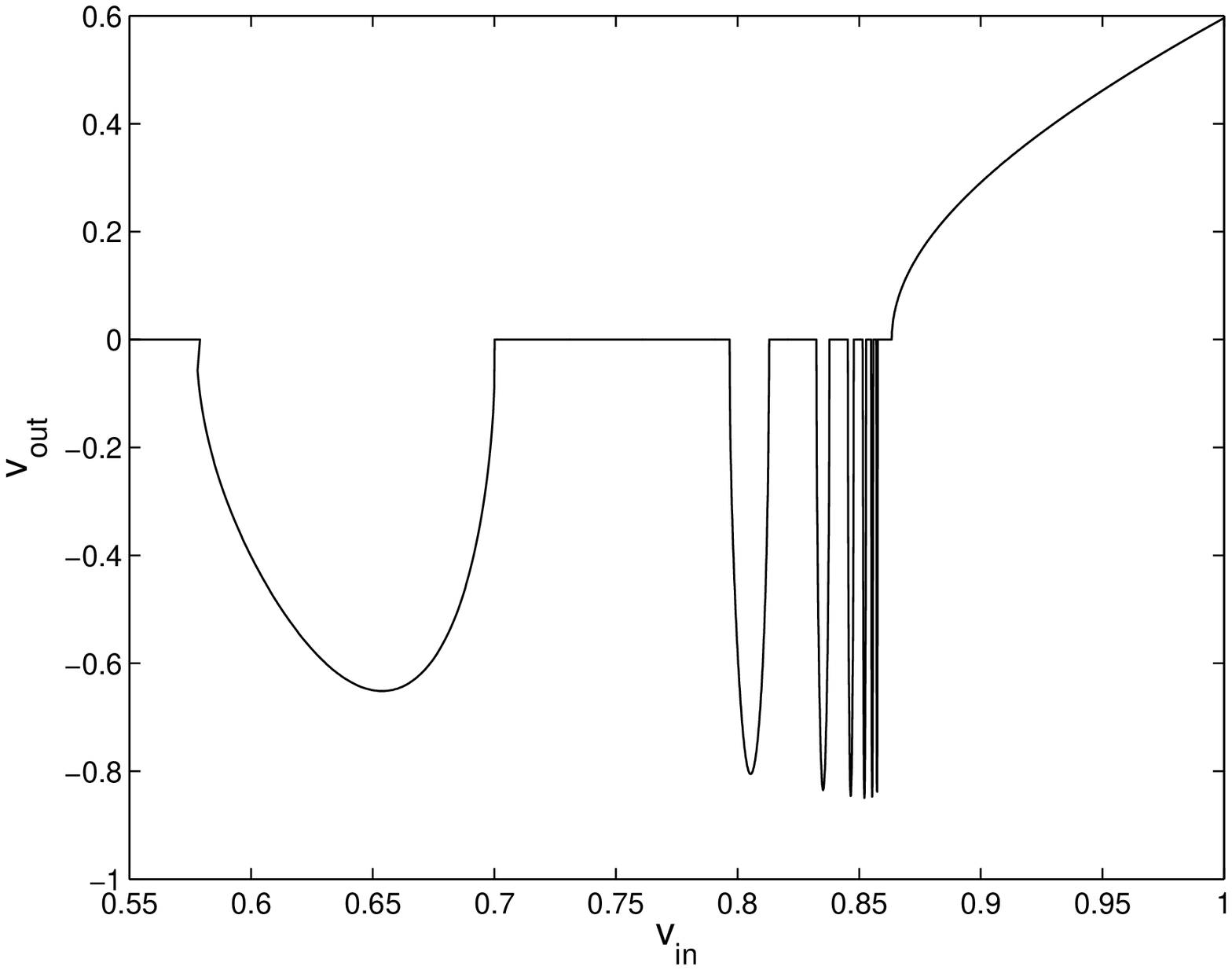}
\caption{Left:  The exit velocity graph for equation~\eqref{eq:TanYang} for $\beta=0.2$, showing reflection windows and a variety of more complex fractal-like structures.  Right: The same figure with all but the main resonant reflections removed.}
\label{fig:R012}
\end{center}
\end{figure}

In figure~\ref{fig:R012}, the exit velocity graph of~\eqref{eq:TanYang} shows that even at $\beta=0.2$, the ODE dynamics display a complex fractal-like structure in addition to the reflection windows, which are not seen in the PDE dynamics for such small values of $\beta$.  The numerical value of the crititcal velocity is $\vc=0.86$, close to the value $\vc=0.936$ found in~\cite{TY:01}.

\begin{figure}
\begin{center}
\includegraphics[width=4in]{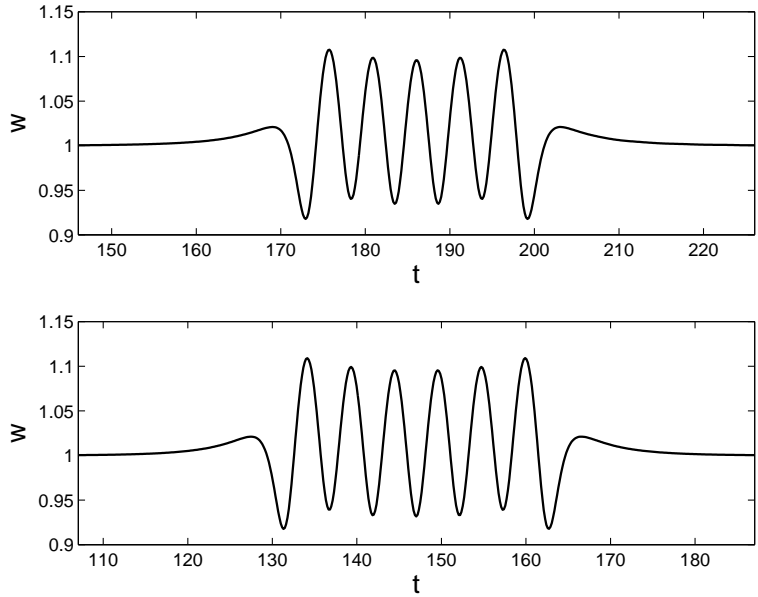}
\caption{Plots of the $w(t)$ component of~\eqref{eq:TanYang} with initial velocity $v= 0.09988$ (top) and $v=0.13464$ (bottom) and $\beta=0.05$, showing the 2-5 and 2-6 resonances.}
\label{fig:BT009}
\end{center}
\end{figure}

The numerical solutions of~\eqref{eq:TanYang} qualitatively explain the resonance windows.  In figure~\ref{fig:BT009}, the $w(t)$ components of the solutions with initial velocity $v$ at the center of the first two resonance windows (actually the points tangents to the line $\vout=-\vin$).  In the leftmost window, the oscillator $w(t)$ is excited, oscillates about 5 times and then is de-excited.  In the next window, $w(t)$ oscillates 6 times.  In each of the successive windows, $w(t)$ oscillates one more time before it is extinguished.  We will refer to the first window as the 2-5 window and the second window as the 2-6 resonance window. Recall that no such windows have been found in the PDE dynamics for this value of $\beta$, but such windows have been found the in the ODE dynamics for all values of $\beta$.   Tan and Yang demonstrated a width oscillation in the PDE solutions in analogy with that shown here.  The minimum value of $n$ in the 2-$n$ resonance decreases with increasing $\beta$.  There does exist a 2-1 resonance with velocity $v=0.649$ in the ODE dynamics shown in figure~\ref{fig:R012}, while the first resonance window found in the PDE simulations is the 2-$2$ resonance at about $v=0.9$ in figure~\ref{fig:tanyang}

\subsection{A further simplified model}

The model~\eqref{eq:TanYang}  bears a striking resemblance to the system derived in~\cite{FKV:92} to study the two-pass resonance in the sine-Gordon equation with defect, and analyzed in~\cite{GH:04}.  In that case, however, the situation is much simpler: the term equivalent to $w$ in~\eqref{eq:TanYang} occurs only linearly, and the potential and coupling terms, equivalent to $F(X/w)$ here, are functions of $X$ alone. This allows us to solve the analog of~\eqref{eq:TY2} by variation of parameters to solve for this term and then insert it into the equivalent of equation~\eqref{eq:TY1}, a critical step in our analysis.  In our numerically-computed solutions displaying the two-bounce resonance for small values of $\beta$, the width $w$ undergoes only a small oscillation about its initial width $w=1$.  Therefore, we may partially linearize system~\eqref{eq:TanYang}, which allows us to proceed in the same manner as we have for the sine-Gordon system.  We find reasonable agreement, with a few notable differences, between the two ODE systems.  We will discuss the linearized theory first and then discuss corrections due to the nonlinearity.

\begin{figure}
\begin{center}
\includegraphics[width=3in]{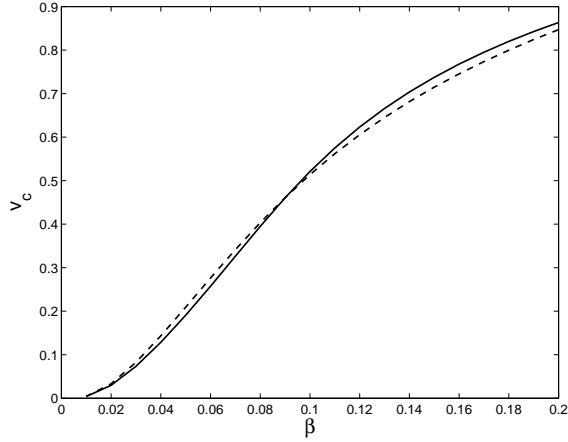}
\caption{The critical velocity for capture as a function of the coupling $\beta$ for the fully nonlinear system~\eqref{eq:TanYang} (solid) and the simplified system~\eqref{eq:simplified} (dashed).}
\label{fig:vc}
\end{center}
\end{figure}

\begin{figure}
\begin{center}
\includegraphics[width=3in,angle=-90]{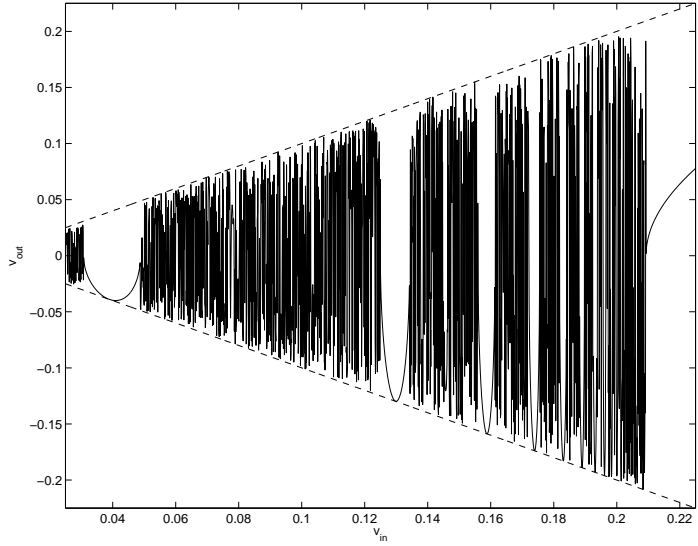}
\caption{The exit velocity graph for the simplified system~\eqref{eq:simplified}, showing qualitative agreement with figure~\ref{fig:R003}.}
\label{fig:SR002}
\end{center}
\end{figure}

Allowing $w=1+W$, where $W$ is considered small, expand all the terms in $W$, and keep only leading-order terms.  We arrive at the reduced system:
\begin{subequations}
\begin{align}
\diffn{X}{t}{2}&= 16\beta\left(F'(X)+G'(X)W\right); \label{eq:simplified_a}\\
\diffn{W}{t}{2}+ \frac{16}{\pi^2}W&=\frac{48\beta}{\pi^2}G(X), \label{eq:simplified_b}
\end{align}
\label{eq:simplified}
\end{subequations}
where
$$
G(X)=-\left(X F(X) \right)'.
$$
Figure~\ref{fig:vc} shows that this simplified equation gives an accurate estimate of the critical velocity for small values of $\beta$ based on numerical simulation.  Figure~\ref{fig:SR002} shows the equivalent of figure~\ref{fig:R003} with the same value of $\beta=0.05$ for the simplified equations.   It shows that the qualitative picture, chaotic scattering interupted by resonance windows for $v<\vc$, is the same, while the actual location of those windows varies greatly.  In the case $\beta=0.05$, the simplified equation has a 2-4 window, while the full equation's first resonance is 2-5.  As $\beta$ was decreased further, the agreement between the two systems improved.
We  rescale time by allowing $t \to 4 \sqrt{\beta} t$, transforming equations~\eqref{eq:TanYang}  to
\begin{subequations}
\begin{align}
\ddot X&= F'(X)+G'(X)W;  \label{eq:scaled_a}\\
\ddot W+\lambda^2 W&=\frac{3}{\pi^2}G(X), \label{eq:scaled_b}
\end{align}
\label{eq:scaled}
\end{subequations}
with fast frequency $\lambda$ given by
\begin{equation}
\label{eq:lambda}
\lambda=\frac{1}{\pi\sqrt{\beta}}.
\end{equation}
The dot notation will be used for derivatives with respect to the scaled time.  The conditions in backward time as $t \to -\infty$ become:
\begin{equation}
X \to -\infty;\; 
\dot X \to \Vin>0;\;
W \to 0 \
\dot W \to 0. \label{eq:init}
\end{equation}
We will  use capital $V$ to represent velocities in the scaled time $t$ and lower-case $v$ for velocities in the physical time.

\section{Determination of energy change and critical velocity}\label{sec:DeltaE}
\label{section:DeltaE}
\subsection{Setup of Melnikov Integral for $\Delta E$}
\begin{figure}
\begin{center}
\includegraphics[width=4in]{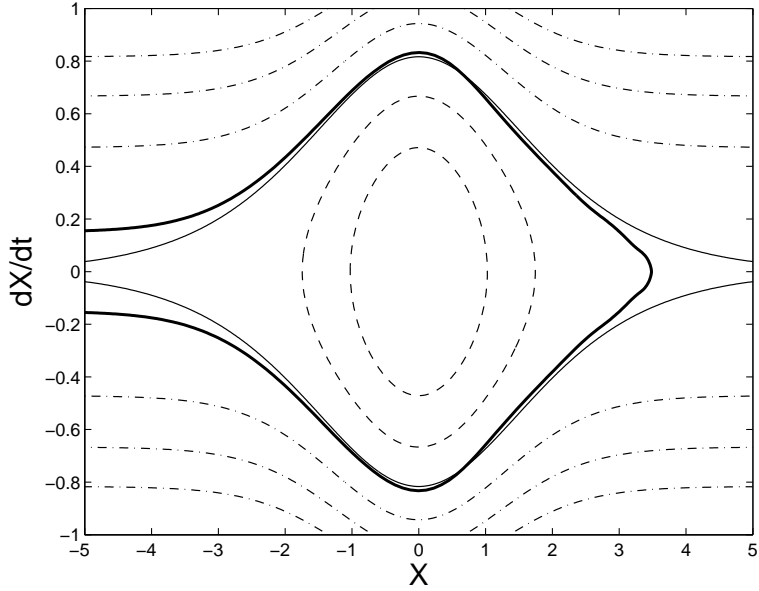}
\caption{The $X$ phase plane, showing trapped (dashed), untrapped (dash-dot), and separatrix (thin solid) orbits, corresponding to level sets of~\eqref{eq:energy}. Superimposed is the $X-\dot X$ projection of the 2-6 resonant solution to the fully nonlinear equations~\eqref{eq:TanYang} with $\beta=0.05$ (thick solid line).}
\label{fig:phasespace}
\end{center}
\end{figure}

First, note that if $W$ is held equal to zero, equation~\eqref{eq:scaled_a} has the phase space shown in figure~\ref{fig:phasespace}, showing three distinct types of orbits: closed orbits, corresponding to a pair of solitons bound together as a breather, unbounded orbits, corresponding to two solitons passing each other by, and orbits heteroclinic to degenerate saddle points at $(X,\dot X)=(\pm \infty,0)$---separatrices---that form a boundary between  the two regimes.  These orbits correspond to level sets, where the energy 
\begin{equation}
\label{eq:energy}
E= \half \dot X^2 - F(X)
\end{equation}
is negative, positive, or zero, respectively. As $W$ is allowed to vary, solutions may cross the separatrices.   We will show below that $W$ remains $O(\sqrt{\beta})$ by variation of parameters~\eqref{eq:varparams} below and, thus, that perturbation methods are applicable.

We wish to asymptotically analyze orbits near the separatrix (see figure~\ref{fig:phasespace}), since two solitons are initially captured when they cross the separatrix and are reflected or transmitted when they cross it a second time and may escape. We first determine the energy loss as a soliton goes from $X=-\infty$ to $X=+\infty$ by computing an energy integral called a Melnikov integral~\cite{M:63}.  A Melnikov integral is a perturbative device for measuring the change of energy in a given system.  It is simply the integral of the time rate-of-change of the energy along some trajectory in the unperturbed problem.  A zero of the Melnikov integral is commonly a necessary condition for chaos in low-dimensional dynamical systems~\cite{GH:83}.  In our case, we simply wish to calculate a change in energy.

The calculation has been simplified significantly from that given in~\cite{GH:04}, in a manner that yields additional insight into the form of the energy loss.  In particular, we do not need to keep track of whether certain functions possess even or odd symmetry, and we find in an elementary way that the change of energy is negative.  First, we note that the separatrix is given by the level set $E=0$, therefore, along the separatrix, equation~\eqref{eq:energy} may be solved for $X(t)$, giving
\begin{equation}
\diff{\Xs}{t}= \sqrt{2 F(\Xs)}.
\label{eq:deltadot}
\end{equation}
Given the function $F$ in~\eqref{eq:F}, it is not possible to find the separatrix orbit $\Xs(t)$ in closed form.  The time-dependent energy exactly satisfies the differential equation
\begin{equation}
\diff{E}{t} = \left(\ddot X - F(X) \right) \dot X =  \dot X G'(X) W= \left( \diff{}{t} G(X(t))\right) W,
\label{eq:DeltaESetup}
\end{equation}
where we have used equations~\eqref{eq:scaled_a}.  We approximate the change in energy for one nearly heteroclinic orbit along the separatrix (from one saddle at infinity to the next saddle approach) by approximating $X(t)$ in~\eqref{eq:DeltaESetup} with the known separatrix solution $\Xs(t)$.  We integrate~\eqref{eq:DeltaESetup} along the length of the orbit and integrate by parts to find the total change in energy:
\begin{equation*}
\begin{split}
\Delta E & = \intinf   \left( \diff{}{t} G(\Xs(t))\right) W\, dt   \\
&= - \intinf G(\Xs(t)) \diff{W(t)}{t}\, dt,
\end{split}
\end{equation*}
where we have integrated by parts.  Given the initial condition~\eqref{eq:init}, with $V=0$ for the separatrix case, we may solve equation~\eqref{eq:simplified_b} for $W(t)$ using variation of parameters:
\begin{equation}
W(t) = \frac{-3}{\pi^2\lambda} \cos{\lambda t} \int_{-\infty}^{t} G(\Xs(\tau)) \sin{\lambda \tau} \, d\tau
+\frac{3}{\pi^2\lambda} \sin{\lambda t} \int_{-\infty}^{t} G(\Xs(\tau)) \cos{\lambda \tau} \, d\tau,
\label{eq:varparams}
\end{equation}
(again approximating $X(t)$ by $\Xs(t)$) and
$$
\diff{W(t)}{t} = \frac{3}{\pi^2} \sin{\lambda t} \int_{-\infty}^{t} G(\Xs(\tau)) \sin{\lambda \tau} \, d\tau
+\frac{3}{\pi^2} \cos{\lambda t} \int_{-\infty}^{t} G(\Xs(\tau)) \cos{\lambda \tau} \, d\tau
$$
Setting 
$I_{\rm s}(t) = \int_{-\infty}^{t} G(\Xs(\tau)) \sin{\lambda \tau} \, d\tau$ and
$I_{\rm c}(t) = \int_{-\infty}^{t} G(\Xs(\tau)) \cos{\lambda \tau} \, d\tau$, we find that 
\begin{equation}
\begin{split}
\Delta E  &= - \frac{3}{\pi^2}\intinf    I_{\rm s}(t)  I_{\rm s}'(t) dt    
- \frac{3}{\pi^2}\intinf    I_{\rm c}(t)  I_{\rm c}'(t) dt   \\
&= -\frac{3}{2\pi^2}(   I_{\rm s}^2(\infty) +  I_{\rm c}^2(\infty)).
\end{split}
\end{equation}
This may be integrated by a standard substitution to yield
\begin{equation}
\begin{split}
\Delta E  &= - \frac{3}{2 \pi^2} \left(
\left(\intinf G(\Xs(\tau)) \sin{\lambda \tau} \ d\tau\right)^2 +
\left(\intinf G(\Xs(\tau)) \cos{\lambda \tau} \ d\tau\right)^2  \right) \\
& =  - \frac{3}{2\pi^2} \left \lvert
\intinf G(\Xs(\tau)) e^{i \lambda \tau} \ d\tau\right \rvert^2.
\end{split}
\label{eq:DeltaEfinal}
\end{equation}
Thus, the problem is reduced to to calculating the integral~\eqref{eq:DeltaEfinal}.  In fact, because in this case $G(\Xs(t))$ is an even function, 
\begin{equation}
\label{eq:DeltaEFinal2}
\Delta E= - \frac{3}{2\pi^2} I_{\rm c}(\infty)^2.
\end{equation}
  Note that this shows the change in energy is generically negative when we assume $W \to 0$ as $t \to -\infty$.  In fact, it must be negative, as the system conserves an energy that is positive-definite as $\abs{X} \to \infty$, and no energy resides in the width oscillation initially.  Using $\Delta E = -\frac{\vc^2}{2}$, we find
$$
\vc= \frac{\sqrt{3}}{\pi} I_{{\rm c},\infty}
$$
where $ I_{{\rm c},\infty}=  I_{\rm c}(\infty)$.  The integral in equation~\eqref{eq:DeltaEFinal2} may be solved numerically by converting it into a differential equation, which may be integrated simultaneously with equation~\eqref{eq:deltadot}.  Alternatively, we derive closed-form approximations to $\Delta E $ $\vc$ in section~\ref{sec:complexvariables} below using complex analysis.
\begin{figure}[htb]
\begin{center}
\includegraphics[width=3in]{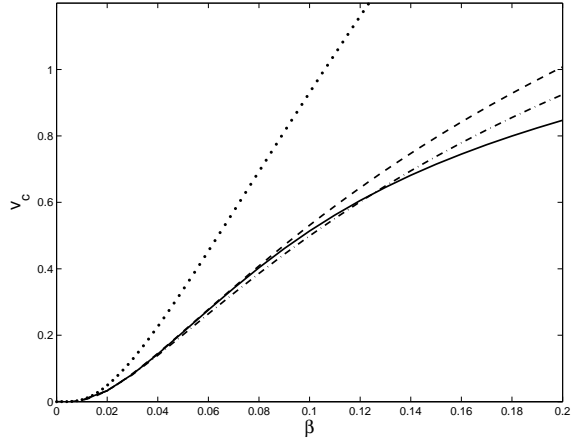}
\caption{The critical velocity of the ODE system~\eqref{eq:simplified} computed via direct numerical simulation (solid line), the first-order asymptotic approximation (dots) and the second-order asymptotic approximation (dashed line) from~\eqref{eq:vseries}, and via numerical evaluation of integral~\eqref{eq:DeltaEfinal}.}  
\label{fig:vseries}
\end{center}
\end{figure}

\subsection{A Generalizaton of the calculation}
Next, we briefly mention two generalizations of the Melnikov calculation above that will be useful later.  First, suppose that instead of approaching zero as $t \to -\infty$, 
$$W \sim \frac{3}{\pi^2\lambda} I_{{\rm c},\infty} \cW \sin{\lambda(t-\phi)}, $$
Then the change of energy will be given by 
$$
\Delta E = \frac{3 I_{{\rm c}, \infty}^2}{\pi^2}\left(-\half + \cW \cos{\lambda\phi}\right).
$$
As $X$ traverses the heteroclinic orbit in the reverse direction, the sign of $G'(\Xs (t))$  in~\eqref{eq:DeltaESetup} is reversed, which leads to:
\begin{equation}
\label{eq:DeltaEreturn}
\Delta E = \frac{3 I_{{\rm c},\infty}^2}{\pi^2}\left(-\half - \cW \cos{\lambda\phi}\right).
\end{equation}
For a resonance to occur, the change of energy calculated in the first Melnikov integral must cancel with the energy jump on the return trip.  Assume the forward heteroclinic orbit has ``symmetry time'' $t_1$ at which $X=0$, with symmetry time $t_2$ on the return trip.  Then, by~\eqref{eq:varparams}, as $t\to\infty$ on the forward heteroclinic orbit, 
\begin{equation}
W(t) \sim \frac{3}{\pi^2\lambda} I_{{\rm c},\infty} \sin{\lambda(t-t_1)}.
\label{eq:Delta_E_general}
\end{equation}
For an exact resonance to occur, the energy change along the two heteroclinics must cancel, leading to the condition
$$
\Delta E_1  + \Delta E_2 =- \frac{3 I_{{\rm c},\infty}^2}{2\pi^2} +
 \frac{3 I_{{\rm c},\infty}^2}{\pi^2}\left(-\half - \cos{\lambda(t_2-t_1)}\right)=0,
$$
obtained by combining~\eqref{eq:DeltaEfinal} and~\eqref{eq:DeltaEreturn} with $\cW=1$ and $\phi=t_2-t_1$. Thus $\cos{\lambda(t_2-t_1)}=-1$, or 
\begin{equation}
\label{eq:t2mt1}
t_2-t_1 = \frac{(2n+1)\pi}{\lambda}.
\end{equation}
This differs from the equivalent resonance condition in~\cite{GH:04}, in which $t_2-t_1 = 2\pi n/\lambda$.  The difference arises because in that system the equivalent term to $G(X)$ was an odd function, whereas here $G$ is even.

Many analyses of two-bounce and two-pass resonance phenomena have been based on the assumption that $t_2-t_1= \frac{2\pi n}{\lambda} + \delta$ for some undetermined $\delta$, a phase shift that accounts for unidentified physical processes that have not been modeled.  Equation~\eqref{eq:t2mt1} shows that in this case $\delta = \frac{\pi}{\lambda}$.  It is worth computing a linear fit of $t_2-t_1$ vs.\ $n$ for comparison with earlier studies and, we will see, for comparison with the analogous result for the fully nonlinear ODE~\eqref{eq:TanYang}.   At $\beta=0.05$, we find the linear fit $t_2-t_1= 4.935\left(n + \half\right) - 0.011$ and for $\beta=0.2$,  we find $t_2-t_1\approx4.931 \left(n + \half\right)+ 0.139$, whereas $\frac{2\pi}{\lambda}\approx4.9348$.  Therefore, we see that to leading order, relation~\eqref{eq:t2mt1} holds, and that the agreement improves with decreasing $\beta$.

\subsection{Evaluation of critical velocity using complex analysis}
\label{sec:complexvariables}

Since $\lambda$ is large and $G(X)$ is analytic, the calculated change of energy~\eqref{eq:DeltaEfinal} is exponentially small.  In a calculation for a similar system, we were able to calculate the analogous integrals explicitly because $\Xs(t)$ was known in a simple closed form~\cite{GH:04}.  In the present case, we are forced instead to expand the integrand of~\eqref{eq:DeltaEfinal} about a certain (branch) pole.  Given the form of the potential~\eqref{eq:F}, $F$ has a pole whenever $\sinh{X}=0$ and the numerator of $F$ is nonzero or has a zero of order less than 3.  The nearest pole to $X=0$ occurs at $X=i\pi$.  Along the separatrix
\begin{equation}
\label{eq:sep}
 \diff{X}{t} = \sqrt{2 F(X)},
 \end{equation}
so we let $t^*$ be chosen such that $X(t^*)=0$ given the initial condition $X(0)=0$.  This gives the formula:
\begin{equation}
\label{eq:tstar}
iT \equiv t^* = \int_{0}^{t^*} dt =   \int_{0}^{i\pi}\sqrt{\frac{1}{2F(X)}} \, dX =
\frac{i}{\sqrt{2}} \int_{0}^{\pi} \sqrt{\frac{\sin^3 y}{\sin y - y\cos y} } dy
 \approx 2.10392\, i.
\end{equation}
We expand~\eqref{eq:sep} about $X=i\pi$ and $t=t^*$ and find
\begin{align*}
\left(\frac{(-1)^{-1/4}}{\sqrt{2\pi}} (X-i\pi)^{3/2} +O((X-i\pi)^{9/2})\right)\,dX& =  \ dt \\
\frac{(-1)^{-1/4}\sqrt{2}}{5\sqrt{\pi}} (X-i\pi)^{5/2} +O((X-i\pi)^{11/2}) & =  t-t^* 
\end{align*}
which may be inverted to form
$$
X-i \pi = (-1)^{1/10} \pi^{1/5}2^{-1/5} 5^{-2/5} (t-t^*)^{2/5} + O((t-t^*)^{8/5}).
$$
Based on the expansion
$$
F(X)=i \pi (X-i\pi)^{-3} + O(1)
$$
we compute the two leading order terms of the integrand of~\eqref{eq:DeltaEfinal}
$$
G(\Xs(t))= \frac{(-1)^{3/5} \pi^{6/5}2^{4/5}3}{5^{8/5}} (t-t^*)^{-8/5} +
 \frac{(-1)^{1/5} \pi^{2/5}2^{8/5}} {5^{6/5}} (t-t^*)^{-6/5} + O( (t-t^*)^{-2/5} ).
 $$
 Therefore~\eqref{eq:DeltaEfinal} involves integrals of the type
 $$
I(\lambda,T,p)= \intinf e^{i \lambda t} (t-i T)^p \, dt
 $$ 
 with $\lambda>0$, $T>0$, and $p<0$. Here $iT$ is the branch pole, from which a branch line extends vertically to $i\infty$.  By a shift of contour and a change of variables to $z=i \lambda(t-iT)$, this can be replaced by an integral over the Hankel contour $\gamma$, which starts at $-\infty$ below the real axis, circles zero once in the positive direction, and returns to $-\infty$ along (and above) the real axis~\cite{CKP}.
 $$
I(\lambda,T,p)= \frac{(-i)^{p+1}}{\lambda^{p+1}} e^{-\lambda T} 
\int_\gamma e^z z^p dz
$$
which forms part of a familiar representation of Euler's gamma function and yields the exponentially small term
\begin{equation}
I(\lambda,T,p)= (-1)^{-p/2} 2 \sin{\left((p+1)\pi\right)} \G{(p+1)} e^{-\lambda T}\lambda^{-(p+1)}.
\label{eq:Intlambda}
\end{equation}

Using~\eqref{eq:Intlambda} and standard trigonometric and gamma function identities, we evaluate the integral in~\eqref{eq:DeltaEfinal} 
\begin{equation}
\intinf G(\Xs(\tau)) e^{i \lambda \tau} \ d\tau =
\left(
\frac{(-1)^{7/5}  2^{4/5} \pi^{6/5}}{5^{3/5}} \Gamma\left(\frac{2}{5}\right)\sin{\frac{2\pi}{5}}\lambda^{3/5}  +
\frac{(-1)^{4/5}  2^{8/5} \pi^{2/5}}{5^{1/5}} \Gamma\left(\frac{4}{5}\right)\sin{\frac{4\pi}{5}}\lambda^{1/5}  + 
O(\lambda^{-3/5}) \right) e^{-\lambda T}.
\end{equation}
As the integrand is real, we choose the branch $(-1)^{1/5} = -1$ above.  Using that $\Delta E =  \frac{\vc^2}{2}$ and the scaling relation given before~\eqref{eq:scaled}, we arrive at the expansion for the critical velocity in physical variables:
\begin{equation}
\vc = \frac{8\sqrt{3}}{5} e^{-T/\pi\sqrt{\beta}}\left(\theta(2/5)\alpha \beta^{1/5} - \theta(4/5)\alpha^2 \beta^{2/5} + \ldots \right),
\label{eq:vseries}
\end{equation}
where $\theta(x)= \sin{\pi x}\G{(x)}$ and $\alpha= \pi^{-2/5} 2^{4/5} 5^{2/5}$. using equations~\eqref{eq:lambda},   \eqref{eq:energy}, as well as the two integrals above.  
Figure~\ref{fig:vseries} shows that the critical velocity is poorly predicted by the first term in this series, but well-predicted up to about $\beta=0.1$ when the second term is added.  The series expansion of the integrand of~\eqref{eq:DeltaEfinal} about $t=t^*$ contains one more integrable  term which does not lead to a visible improvement of the approximation to $\vc$.   In order to improve the approximation, one would have to calculate expansions about the additional singularities of $G(X_S(t))$ further off the imaginary $t$-axis.

\section{Matched asymptotic construction of solutions}
\label{sec:matching}
\subsection{The expansion framework}
If $\Vin >\Vc$, then $\dot X$ remains positive for all time and $X \to +\infty$ monotonically.  We can call this a one-pass transmitted solution.  A ``pass'' will occur each time $X=0$, when the two solitons pass each other and energy is transferred between the translation and vibration modes.  If $\Vin < \Vc$, then the energy is negative after one pass, and the solitons reverse direction, setting up the second pass.  On the first pass, the change of energy was shown in~\eqref{eq:DeltaEfinal} to be negative, but on subsequent passes, it may take either sign, by~\eqref{eq:DeltaEreturn}.  On the second, and subsequent, passes the solitons may escape if the energy is positive, or may be reversed again.  We will focus primarily on the case that the solitons interact twice before escaping.

Following~\cite{GH:04}, we construct 2-pass solutions by a matched asymptotic expansion.  The solution consists of sequences of nearly heteroclinic orbits connected to near saddle approaches at $X=\pm\infty$.  The change in energy from one saddle approach to the next is approximated by the Melnikov integral calculated in section~\ref{sec:DeltaE}.
The 2-bounce solution can be constructed from the following 5 pieces:
\begin{enumerate}
\item A near saddle approach to $X=-\infty$ with energy $E_0=\half \Vin^2$, such that $\dot X \to \Vin < \Vc$,  as $t \searrow -\infty $ ;
\item a heteroclinic orbit with $dX/dt>0$ such that $X(t_1)=0$, with energy change $\Delta E_1$ given by~\eqref{eq:DeltaEFinal2};
\item a near saddle approach to $X=+\infty$ with negative energy $E=-\half M^2$, such that $X$ achieves its maximum at $t=t^{*}$;
\item a heteroclinic orbit with $dX/dt <0$ such that $X(t_2)=0$,  with energy change $\Delta E_2$ given by~\eqref{eq:DeltaEreturn};
\item and a near saddle approach to $X=-\infty$ with positive energy $E=\half \Vout^2$,  such $\dot X \to -\Vout$, as $t \nearrow \infty$.
\end{enumerate}
The times $t_1$, $t_2$, and $t^*$, as well as the energy levels, remain to be determined below.  In the language of matched asymptotics, the approximations at steps 1, 3, and 5 are the ``outer solutions'' and steps 2 and 4 are the ``inner solutions.''

A comment about the last step is in order.  For general initial velocity $\Vin$, the energy at step 5 will not match the energy at step 1.  If these two energies match exactly, then we say the solution is a 2-pass resonance.  If the energy at step 5 is positive but less than $\Vin^2/2$, then the solution is in the 2-pass window, and may be called an incomplete resonance.  Physically, the solitons reflect off each other, but with reduced speed and with significant energy remaining in their width oscillation.  The outer edges of the window will be given by velocities where the energy at step 5 is identically zero.  This defines the width of the windows. If at step 5, the energy is instead negative, then the solution remains trapped for another step, alternating between negative energy near-saddle approaches to $X=-\infty$ and $X=\infty$ until enough energy is returned to $X$ such that $E=\Vout^2/2$, and $X\to \pm \infty$.  Non-resonant solutions and higher resonances are explained in section~\ref{sec:generalized}.

For the simpler sine-Gordon system, we wrote down a general asymptotic formula for $n$-pass solutions, calculated the location of 3-pass windows, and calculated the widths of the 2-bounce windows~\cite{GH:04}.  Analogous results are possible in the present situation and are discussed below in section~\ref{sec:generalized}, although in less detail than in the previous paper.

We use the method of matched asymptotic expansions, as in~\cite{DH:02,DH:03, GH:04}.  The heteroclinic orbits along the separatrix are matched (forward and backward in time) to the finite time singularities associated with the near-saddle approaches. We will not make use of the two positive energy expansions, so we will not compute them.  They enter the analysis  when the energy change calculated over the heteroclinic orbits in the above section will then be used to connect the positive and negative energy expansions. 

\subsection{Asymptotic description of heteroclinic orbit for large $X$}

We first construct an expansion of the ``inner solution,'' given by the heteroclinic orbit. Along the heteroclinic orbit, ${\dot X}^2/2 = F(X)$.  Setting $X=0$ at $t=t_1$, the trajectory is given as the solution to
\begin{equation}
\label{eq:zero_energy}
\int_{t_1}^t dt' = \int_0^X \frac{dX'}{\sqrt{2F(X')}} = \int_0^{X_0} \frac{dX'}{\sqrt{2F(X')}} z+ \int_{X_0}^X \frac{dX'}{\sqrt{2F(X')}} 
\end{equation}
for an arbitrary $O(1)$ constant $X_0<X$.  The first integral is $O(1)$ and will be asymptotically dominated by the second.  For $X\gg 1$, we may approximate the potential by 
\begin{equation}
\label{eq:Fasympt}
F(X) \sim  4 (X-1) e^{-2X}.
\end{equation}
If we let $Z=X-1$, then 
\begin{equation}
\label{eq:XtoZ}
F(X) \sim  (4 e^{-2}) Z e^{-2Z},
\end{equation}
which we substitute into~\eqref{eq:zero_energy} and get
$$
t-t_1= \frac{e}{\sqrt{8}}  \int  \frac{e^{Z}}{\sqrt{Z}} dZ +O(1).
$$
We make the substitution $Z=C+y$, where $C\gg 1$ and $y=O(1)$, as motivated in the next section, and expand the integral in powers of $C^{-1}$, yielding
\begin{align}
\label{eq:expand_heterocline}
t-t_1=\frac{e^{1+C}}{\sqrt{8C}}\left(e^y + \frac{1}{2C} (1-y)e^y + O\left(\frac{1}{C^2}\right) \right) + O(1)
\end{align}

\subsection{Asymptotic description of the saddle approach near $X= \infty$}
As the width perturbation $W(t)$ remains small, we may construct approximate solutions from solutions to equation~\eqref{eq:scaled_a} with $W=0$.  Since $F(X)\to 0$ as $\abs{X}\to\infty$, $(\infty,0)$ is a degenerate fixed point and is of saddle-type.
First we compute the near-saddle approach at $X=+\infty$, under the approximation that the solution has small constant energy given by $$ E = -\frac{M^2}{2}$$
with $M \ll 1$. 
So that, using~\eqref{eq:XtoZ}, the near-saddle expansion satisfies:
$$
\half \dot Z^2 - 4 e^{-2} Z e^{-2Z} = -\frac{M^2}{2}.
$$
We make the expansion $Z= C + y$, where $C \gg 1$ is determined in~\eqref{Cn} and show $y=O(1)$. Then 
$$
\half \dot y^2 - 4 e^{-2}(C+y) e^{-2C}e^{-2y} = -\frac{M^2}{2}.
 $$
We define $C$ by
\begin{equation}
\label{Cn}
8 C e^{-2}e^{-2  C} = M^2
\end{equation}
and let $T = M t$, which gives the simplified equation
\begin{equation}
\label{eq:scaledNSA}
\left(\frac{dy}{dT}\right)^2 - \left(1+\frac{y}{C}\right)e^{-2y} = -1.
\end{equation}

\subsection{The matching procedure for near and exact 2-pass resonances}
\subsubsection*{The simplest asymptotic approximation}
Ignoring the term $\frac{y}{C}$ in~\eqref{eq:scaledNSA}, as $C\gg1$, the near saddle approach takes the form
$$
e^y = \cos{(T-T^{*})}
$$
where $T^{*}$ is the ``center time'' at which the near saddle approach comes
closest to the degenerate saddle at $(X,\dot X)=(\infty, 0)$.  This has
finite-time singularities forward and backward in time. Backwards in time, this is singular as $T-T^{*} \searrow -\frac{\pi}{2}$, and
may be asymptotically expanded as 
\begin{equation}
\label{eq:expansion}
e^y = T-T^{*} + \frac{\pi}{2}.
\end{equation}
For large $X$, the  heteroclinic orbit in~\eqref{eq:expand_heterocline} is
asymptotically approximated using~\eqref{Cn}, to leading order in $C^{-1}$, by 
 $$
e^y = T-T_1 
$$
with a similar expression for $T_2-T$ along the return heteroclinic.  The algebraic growth of the heteroclinic orbit matches to the finite time singularities of the near-saddle approach only if
$$
T^{*}-T_1 = \frac{\pi}{2},
$$
which, combined with a similar relation for $T_2-T^{*}$, yields
$T_2-T_1 = \pi$ as in our analysis of the sine-Gordon model, or, in the unscaled time variable
\begin{equation}
\label{t2-t1_a}
t_2 - t_1  = \frac{\pi}{M},
\end{equation}
which shows the energy dependence of the period.  The energies at steps 1 and 3 are related by 
$\frac{\Vin^2}{2} + \Delta E_1 = -\frac{M^2}{2}.$

\subsubsection*{Exact Resonance Condition:} 
To this point, the calculation has been valid for general $\Vin<\Vc$.  We now specialize to the case of exact 2-pass resonance. In a resonant solution the second energy jump must balance the first $\Delta E_2 = -\Delta E_1$, a condition for which is given in~\eqref{eq:t2mt1}, implying $\lambda/(2n+1)$.  Thus the resonant initial velocity $V_n$ solves
\begin{equation}
\label{eq:VnDeltaEMn}
\frac{V_n^2}{2} + \Delta E_1 = -\frac{M_n^2}{2}.
\end{equation}
Solving this for $V_n$, using that $\Delta E_1 = -\Vc^2/2$,
\begin{equation}
v_n=\sqrt{\vc^2 - \frac{16}{\pi^2 (2n+1)^2}},
\label{eq:vn_comp1}
\end{equation}
where scaling~\eqref{eq:lambda} has been used to convert this result back to the physical variables.
Also $n_{\rm min}(\beta)$ is given by the smallest integer $n$ that makes $v_n$ a real number in equation~\eqref{eq:vn_comp1}. In figure~\ref{fig:vn_computed}, we see that this does a relatively poor job at predicting resonant velocities.

\subsubsection*{An Improved Approximation for Resonance}
The above calculation showed clearly the procedure used to find the resonant velocities.  Here it is refined slightly to improve its accuracy.  After some rearrangement, equation~\eqref{eq:scaledNSA} becomes
\begin{equation}
\label{eq:period_integral}
\frac{dy}{\sqrt{\left(1+\frac{y}{C_n}\right)e^{-2y} -1}}=dt
\end{equation}
Scaling the time variable as before, we  expand this integral in powers of $\frac{1}{C_n}$ and keep the first two terms. This gives 
\begin{equation}
\label{eq:scaledNSA_2}
\frac{dy}{\sqrt{e^{-2y}-1} }+ \frac{1}{2 C_n}\frac{y e^{-2y} dy}{(-1+e^{-2y})^{3/2}} = dT
\end{equation}
for the near saddle approach, which has solution:
\begin{equation}
\label{scaled_NSA_2_sol}
e^y - \sin{\left( \frac{1}{2 C_n +1} \frac{y}{\sqrt{e^{-2y}-1}} \right)} 
= \cos{\left(\frac{T-T^{*}}{1+ \frac{1}{2C_n}}\right)}.
\end{equation}
As  $T-T^{*} \searrow \left(1+ \frac{1}{2C_n}\right) \frac{\pi}{2}$ on the right-hand side, we find $y\to-\infty$.  
Rearranging the expansion of the heteroclinic orbit, we find~\eqref{eq:expand_heterocline}
$$
\left(1-\frac{1}{2 C_n +1}y\right) e^y= \frac{T-T_1}{1+ \frac{1}{2{C_n}}}.
$$
Matching these two approximations gives $T_2-T_1 = \left(1+\frac{1}{2 C_n}\right)\pi$ or
\begin{equation}
\label{t2-t1_b}
t_2 - t_1  = \frac{\left(1+\frac{1}{2C_n}\right)\pi}{M_n},
\end{equation}
a more accurate approximation to the period.  Combining this with~\eqref{eq:t2mt1} yields:
$$
 \frac{\left(1+\frac{1}{2C_n}\right)\pi}{M_n} = \frac{(2n+1)\pi}{\lambda}.
$$
This is still not a closed equation as $C_n$ has yet to be specified.  We may eliminate $M_n$ from this equation and~\eqref{Cn} to obtain an implicit relation that defines $C_n$:
\begin{equation}
\label{Cn_implicit}
32\pi^2 \beta (2n+1)^2 C_n^3 - (2C_n+1)^2 e^{2C_n+2} = 0.
\end{equation}
This has exactly two positive roots as long as $2\pi^2e^{-2}\beta(2n+1)^2 > \frac{3+2\sqrt{3}}{9}e^{\sqrt{3}}$, with the larger root relevant. Thus, we come to the revised estimate of the resonant velocities
\begin{equation}
\label{eq:vn_comp2}
v_n=\sqrt{\vc^2 - \frac{16\left(1+\frac{1}{2C_n}\right)}{\pi^2 (2n+1)^2}}.
\end{equation}
Figure~\ref{fig:vn_computed} shows that this does better than our first estimate.  In a similar computation, we found that this analysis in a neighborhood of $X=\infty$ was enough to determine the resonant velocities.  We find in the next section that we can do better with a numerical criterion based on~\eqref{eq:t2mt1}.

\subsubsection*{A Numerical Condition for Resonance}
In all situations where heteroclinic or homoclinic orbits are matched to near-saddle approaches $t_2-t_1$ equals half the period of~\eqref{eq:scaled_a} with $W$ set to zero,
\begin{equation}
\label{eq:XnoW}
\ddot X - F(X) =0.
\end{equation}  
Let $P(E)$ be the period of the closed orbit of~\eqref{eq:XnoW} with energy $E<0$.  We may solve for $P(E)$ by evaluating the definite integral
\begin{equation}
\label{eq:Period}
P = 2 \int_{-\Xmax(E)}^{\Xmax(E)} \frac{dX}{\sqrt{2}\sqrt{F(X)+E}}
\end{equation}
where $\Xmax(E)$ is the positive root of $F(X)+E=0$. The period $P$ cannot be computed in closed form given the particular potential $-F(X)$ in this problem.  Alternately, one can compute the period simply by integrating equation~\eqref{eq:XnoW} with initial conditions $X=\Xmax(E)$, $\dot X = 0$ until reaching the termination condition $X=-\Xmax(E)$ at the time $P(E)/2$.

\begin{figure}
\begin{center}
\includegraphics[width=4in]{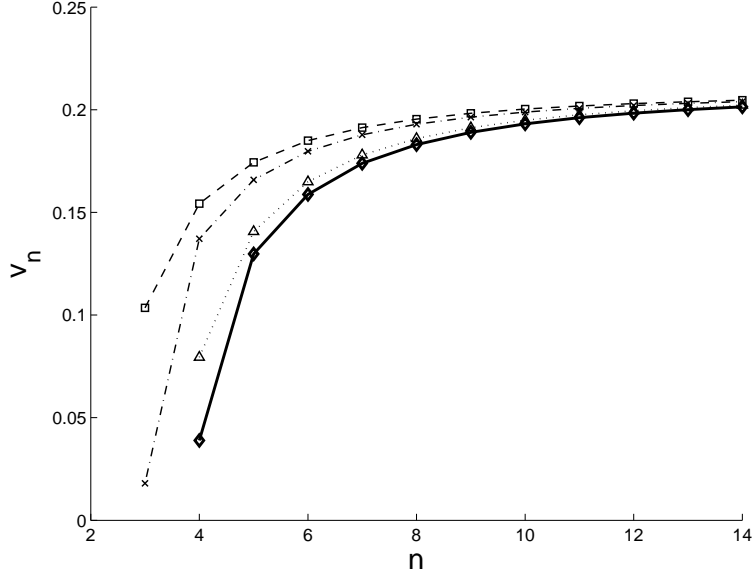}
\caption{The resonant velocities, indexed by the number of complete oscillations of $w(t)$, with $\beta=0.05$.  The thick solid curve at the bottom is the result of direct numerical simulation.  From top to bottom, the other curves are the asymptotic results of equations~\eqref{eq:vn_comp1} and~\eqref{eq:vn_comp2}, and the value involving numerical calculation of the energy level, given the resonant period~\eqref{eq:nlresonance}}
\label{fig:vn_computed}
\end{center}
\end{figure}

In either case, the above calculation must be inverted numerically to yield the energy as a function of the period, using the secant method or some variant.  In the scaled variables, we have, from equation~\eqref{eq:t2mt1}, 
\begin{equation}
\frac{P(E_n)}{2} = \frac{(2n+1)\pi}{\lambda}
\label{eq:nlresonance}
\end{equation}
This is solved for $E_n = -\frac{M_n^2}{2}$, and the resonant velocity is found using~\eqref{eq:VnDeltaEMn}.  This, and the other two approximations are shown in figure~\ref{fig:vn_computed} 
All solutions with positive $n$ up to $n=14$ are shown in the figure.  Approximations~\eqref{eq:vn_comp1} and~\eqref{eq:vn_comp2} both predict the existence of a 2-3 window, while the numerical calculation~\eqref{eq:nlresonance} does not, and no such window is found by direct numerical simulation.

\subsection{Generalizaton to near-resonances and higher resonances}
\label{sec:generalized}
The 2-pass resonant solutions are a countable, and thus measure-zero, family of initial conditions.  Each 2-pass window has finite width whose left and right edges can be found by imposing the conditions that $\Delta E_2 = M^2/2$, so that the output energy is identically zero.  It can be shown that the window widths scale as $n^{-3}$ for large $n$.

In between the 2-pass windows there is a complicated structure consisting of many narrower windows.  These include 3-pass windows, which can be found as follows.  A three-pass resonant solution has three energy jumps.  Just as $W(t)$ and $X(t)$ are even functions about $t^*$ in two-pass solutions, in 3-pass solutions, $W(t)$ and $X(t)$ are odd functions about their center time.  We can place the three ``center times'' at $t=-t_0$, $t=0$, and $t=t_0$, and notice that if the solution is odd, then $\Delta E=0$ at $t=0$.   The change of energy at the second jump is $\Delta E = -3 I_{\rm{c},\infty}^2/\pi^2 (\frac{1}{2} + \lambda t_0)$, which implies 
$$
t_0 = \frac{\left(2n+1 \pm \frac{1}{3}\right) \pi}{\lambda}.
$$
and gives 3-pass resonant solutions with 
$$
v_{3,n\pm}  = \sqrt{\vc^2 - \frac{16}{\pi^2 (2n+1\pm\frac{1}{3})^2}}.
$$
A corrected formula, as in equation~\eqref{eq:vn_comp2}, and a more accurate numerical condition, as in equation~\eqref{eq:nlresonance}, may also be derived.  A general formula for the locations of higher complete resonances can be derived as in~\cite{GH:04}, but this equation must be solved numerically. 

\section{The effect of coupling to a weakly nonlinear oscillator}
\label{sec:nonlinearity}
\begin{figure}
\begin{center}
\includegraphics[width=3in]{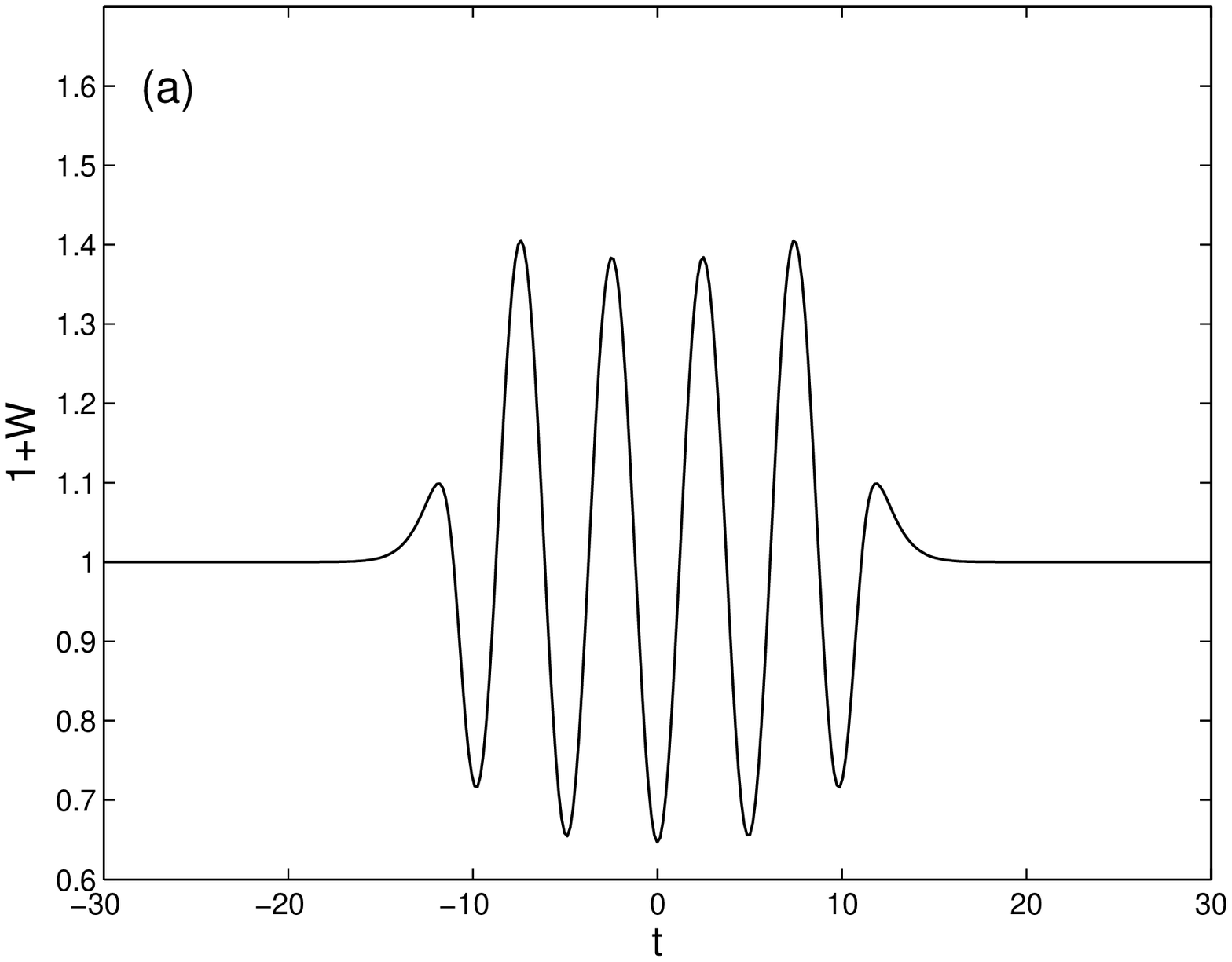}
\includegraphics[width=3in]{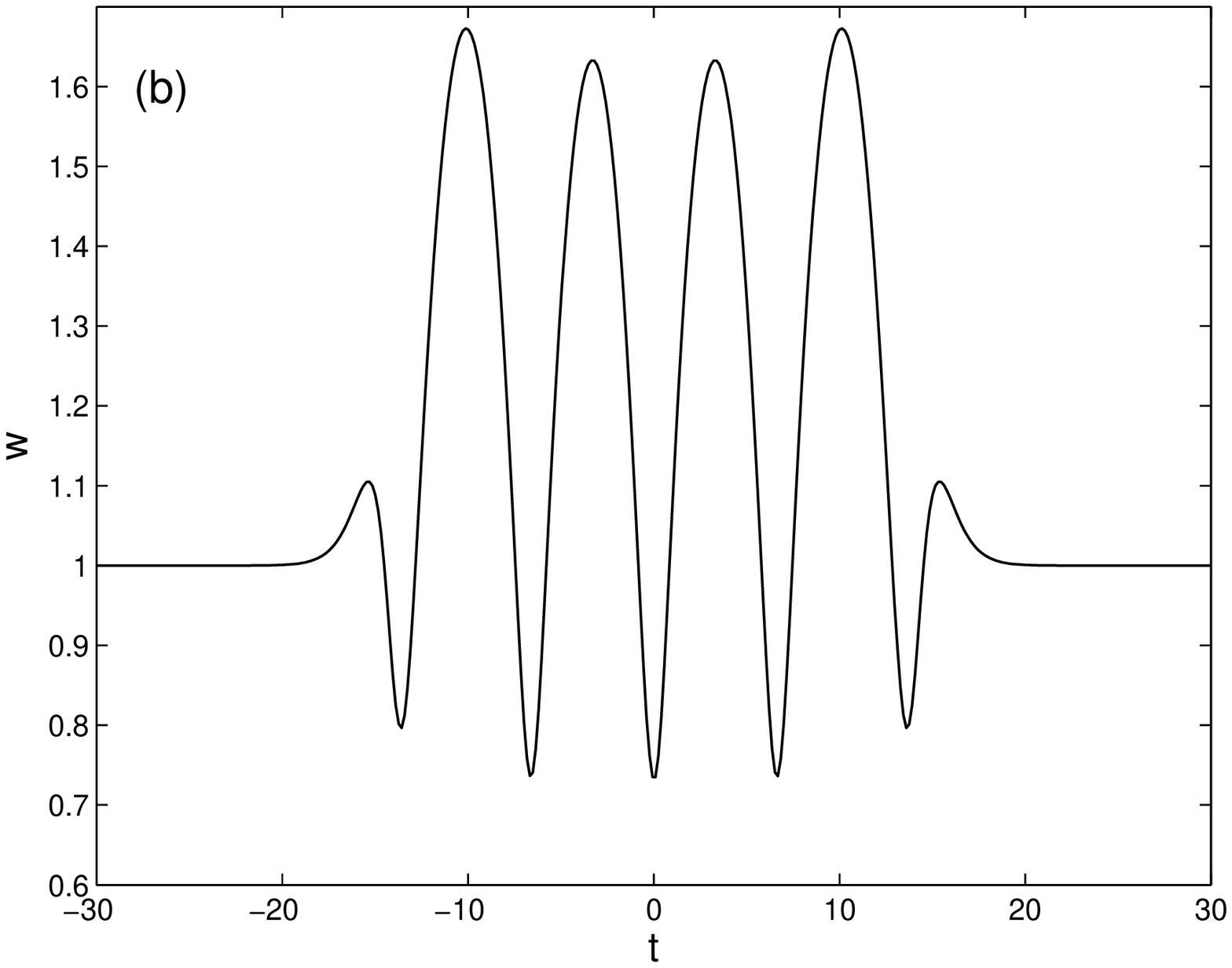}
\caption{The solutions components of the 2-5 resonance of \textbf{(a)} $1+W(t)$ of the simplified equation~\eqref{eq:simplified} and \textbf{(b)} $w(t)$ of the full ODE~\eqref{eq:TanYang}.}
\label{fig:w_vs_W}
\end{center}
\end{figure}

We briefly discuss the discrepancies between the full ODE model~\eqref{eq:TanYang} and the simplified model~\eqref{eq:simplified}, in order to account for the marked difference between the window locations between figures~\ref{fig:R003} and~\ref{fig:SR002}.  The most obvious, seen by comparing the two graphs of figure~\ref{fig:w_vs_W}, where the component $w(t)$ and $1+W(t)$ are both plotted for the 2-5 resonance with $\beta=0.2$.  It is clear that $w(t)$, the fully nonlinear oscillation, has a larger amplitude, a larger period, and the mean about which it oscillates is displaced upward.  As done following equation~\eqref{eq:t2mt1}, we fit $t_2-t_1$ from our numerical simulations and find $t_2-t_1 \approx 5.00\left(n+\half\right)+0.541$ when $\beta=0.05$.  For the case $\beta=0.2$, careful examination shows that $t_2-t_1$ is not approximated that well by a linear fit, with the growth in $t_2-t_1$ slowing as $n$ increased, and $t_2-t_1\approx 6.38\left(n+\half\right) + 0.327$ when the first 10 resonances are used, and $t_2-t_1\approx 6.20 \left(n+\half\right)+ 1.93$ when resonances 11 through 20 are used, and the error in this fit is much larger than in the simplified model, especially when the leftmost windows are included.  We see then that, in addition to a large correction to the oscillation frequency, a significant phase shift appears in the fully nonlinear dynamics.

This shift in amplitude, frequency, and mean value can be explained using a strained coordinate or Poincar\'e-Lindstedt expansion~\cite{KC:96}.  We assume that $w=W+1$ satisfies the homogeneous part of~\eqref{eq:TY2}
\begin{equation}
\label{eq:nonlin}
\diffn{w}{t}{2}=\frac{\lambda^2}{w^2}\left(\frac{1}{w}-1\right)
\end{equation}
where we have used scaling~\eqref{eq:lambda} and have again set $K=1$.  Expanding this as a power series in $W$, we find
\begin{equation}
\label{eq:W_expansion}
\diffn{W}{t}{2} +\lambda^2 W = \lambda^2\left(3 W^2 - 6 W^3 + O(W^4)\right)
\end{equation}
 We look for periodic solutions with initial conditions
 \begin{equation}
 W(0)=\epsilon \text{ and } \dot W (0)=0,
 \label{eq:W_ic}
 \end{equation}
 noting that these initial conditions are somewhat arbitrary, and leaving $\epsilon$ positive and small, but for now undefined.  We expand both the function $W$ and the frequency of oscillations $\Omega$ in powers of $\epsilon$
\begin{equation}
\label{eq:W_PL}
\begin{split}
W &= \sum_{k=1}^{\infty} \epsilon^k W_{k-1}(T) \\
\Omega &=  \sum_{k=0}^{\infty} \epsilon^k \Omega_k 
\end{split}
\end{equation}
where $T=\Omega t$, and assume that the solution has period $2\pi$ in $T$. The equation is satisfied at  each order in $\epsilon$, with the $\Omega_k$ chosen to suppress secular growth terms.  We find that 
$$
\Omega_0 = \lambda ; \; \Omega_1=0;\;  \Omega_2=-\frac{3}{2}\lambda
$$
and
\begin{equation}
\label{eq:W_k}
\begin{split}
W_0 & = \cos{\Omega t}\\
W_1 & = \frac{3}{2} - \cos {\Omega t} -\half \cos{2\Omega t} \\
W_2 &=  -3 + \frac{13}{8} \cos{\Omega t} +  \cos{2\Omega t} + \frac{3}{8}\cos{3\Omega t} 
\end{split}
\end{equation}
Thus the period of oscillation is decreased at larger amplitudes, as found from the least squares fits.

It remains to determine a suitable value of $\epsilon$ in~\eqref{eq:W_ic} and its effect on the resonance.  The full ODE model~\eqref{eq:TanYang} conserves the Hamiltonian 
\begin{equation}
\label{eq:NLEnergy}
H = \half{\dot X}^2 +\frac{2\pi^2}{3} \dot{w}^2  + \frac{8}{3}\left(1+ \frac{1}{w^2}-\frac{2}{w}\right) 
- \frac {16\beta}{w} F\left(\frac{X}{w}\right),
\end{equation}
not written here in canonical variables.
We see from figure~\ref{fig:vc} that $\vc$, and hence $\Delta E$ is approximately the same for the full and simplified ODE systems.  Expanding this in powers of $W=w-1$, we obtain the approximate Hamiltonian
\begin{equation}
\label{eq:Energy_expansion}
H \approx  \half{\dot X}^2 +\frac{2\pi^2}{3} \dot{W}^2  + \frac{8}{3}\left( W^2 - 2W^3 + 3 W^4+\ldots\right) 
- 16\beta F(X) - 16\beta G(X)W + \ldots.  
\end{equation}

As $t\to -\infty$, all of the energy is stored as kinetic energy in the soliton modes $H=\half v_0^2$.  At the symmetry time, which we can set to $t^{*}=0$, $\dot X=0$ and $\dot w=0$, and $X=\Xmax$ is given, if $G(X) W$ is small enough to be ignored, by the solution to $\half v_0^2 - \half \vc^2  = -16 \beta F(\Xmax)$.
Plugging this back into the energy~\eqref{eq:NLEnergy} and using the expansion~\eqref{eq:Energy_expansion} only for the coupling $F(X/w)$ term, we obtain
$$
\half \vc^2 \approx  \frac{8}{3}\left(1+ \frac{1}{(1+W)^2}-\frac{2}{1+W}\right) - 16 \beta G(\Xmax)W.
$$
For resonant velocities sufficiently close to $\vc$ or for small enough $\beta$, $16\beta G(\Xmax)$ is negligibly small, and we can solve the resulting quadratic equation for $W(0)$ and obtain  
$$
W(0)= \frac {\pm \frac{\sqrt 3}{4}\vc}{1\pm \frac{\sqrt 3}{4}\vc}\approx\pm \frac{\sqrt 3}{4} \vc.
$$
For larger values of $16\beta G(\Xmax)$, the equation is cubic in $W$ and the roots may be found by a perturbation expansion around the previously found roots.  We use this value of $W(0)$ as our value of $\epsilon$.  For $\beta=0.05$, it is sufficient to use $\epsilon=\sqrt{3}\vc/4$, which gives a period $5.00$, as was found from the linear fit.  For $\beta=0.2$, we find that weakly nonlinear theory is not useful as the first several terms of the expansion of the Hamiltonian in $W$ of~\eqref{eq:Energy_expansion} are all found numerically to about the same order, so that ignoring the $W^4$ and $W^5$ term in the Poincar\'e-Lindstedt expansion is invalid.

\section{Conclusion}
\label{sec:conclusion}
We have explained many of the phenomena seen in the collision of vector solitons in CNLS.  First we derived and justified a simplified version of a model derived by Ueda and Kath. Using a Melnikov integral, we estimated the critical velocity, and using matched asymptotic expansions near separatrices, we explained how to connect subsequent passes to construct an approximate solution.  Imposing the condition that the total energy change after two passes is zero allowed us to find the locations of the exact two-pass resonant velocities, the centers of the two-pass windows.  It remains to be seen if this phenomenon can be produced in physical experiments, but the experimental setup would appear to be fairly simple.

More importantly, we have elucidated the mechanism underlying two-pass and two-bounce resonance phenomena in general.  The important elements of an ODE model are the following:
\begin{itemize}
\item a ``position'' mode $Z(t)$ that moves in a potential well $V(Z)$, which is localized near $Z=0$, so that the force approaches zero at large distances,
\item a secondary oscillator mode $W(t)$ that acts as a temporary energy reservoir;
\item a term $C(Z,W)$ that couples the two modes together, also localized near $Z=0$, so that coupling decays at large distances.
\end{itemize}
Trapping takes place on the initial interaction if enough energy leaks from the position mode ($Z$) to the energy reservoir ($W$). In that case, the first mode crosses a separatrix curve in its unperturbed phase space.  The  energy change on subsequent interactions may be positive or negative, depending sensitively on the phase of $W(t)$ at the interaction time, even though $W$ may remain exponentially small. Eventually, in any Hamiltonian model, enough energy will eventually be transferred back to the position mode that it returns to the unbounded portions of phase space and escapes.  It is transmitted if it escapes to $+\infty$ and reflected if it escapes to $-\infty$.  All the models we have seen are Hamiltonian, but it should be possible to carry through much of the analysis in the presence of a simple dissipative term.

We believe this mechanism to be present in all the systems which have displayed two-bounce resonance phenomena, including the foundational papers on kink-antikink interactions in nonlinear Klein-Gordon equations~\cite{CP:86,CPS:86,CSW:83,PC:83}.  A finite dimensional model for kink-antikink interactions in the $\phi^4$ model is presented by Anninos et al.~\cite{AOM:91}.  The model they derive is essentially of the form described, but the potential and coupling terms are much more complicated than~\eqref{eq:TanYang}, and it would be quite difficult even to find expansions about poles in the solution, as was done in section~\ref{sec:DeltaE} of the current papers.  What's more there is no natural small parameter measuring the coupling and the difference in time scales of the two modes, as we have here.  The topology is slightly different in the $\phi^4$ kink-antikink problem: the separatrix is given by an orbit homoclinic to infinity, rather than heteroclinic.  This is indeed why the interaction is a ``bounce'' rather than a ``pass.''  We have developed a simple model that displays the same topology as in~\cite{AOM:91} and are in the process of analyzing it as a next step in understanding the two-bounce case.

\section*{Acknowledgements}
We thank Jianke Yang for useful discussions and use of his figures.  Roy Goodman was supported by NSF-DMS 0204881 and by an NJIT SBR grant.


\end{document}